\documentstyle[12pt,psfig,epsf,html]{article}
\def\reference{\parskip 0pt\par\noindent\hangindent 0.5 truecm}
\def\kms{km ${\rm s}^{-1}$}

\newcommand{\micron}{\mbox{$\,\mu$m}}

\begin{document}
%
%
\title{\bf Molecular Hydrogen in the Lagoon: \\ 
H$_2$ line emission from Messier 8}
%


\author{Michael G. Burton$^{1,2}$}


\date{}
\maketitle

{\center $^1$ School of Physics, University of New South Wales,
Sydney, NSW 2052 \\ $^2$ School of Cosmic Physics, Dublin Institute
for Advanced Studies, 5 Merrion Square, Dublin 2, Ireland \\ ~ \\
M.Burton@unsw.edu.au\\[3mm]

}

%
\begin{abstract}
The 2.12\micron\ v=1--0 S(1) line of molecular hydrogen has been
imaged in the Hourglass region of M8\@. The line is emitted from a
roughly bipolar region, centred around the O7 star Herschel 36.  The
peak H$_2$ 1--0 S(1) line intensity is $\rm 8.2 \times 10^{-15} \ erg
\ s^{-1} \ cm^{-2} \ arcsec^{-2}$.  The line centre emission velocity
varies from $-25$\,\kms\ in the SE lobe to $+45$\kms\ in the NW lobe.
The distribution is similar to that of the CO J=3--2 line.  The H$_2$
line appears to be shock-excited when a bipolar outflow from Herschel
36 interacts with the ambient molecular cloud.  The total luminosity
of all H$_2$ lines is estimated to be $\rm \sim 16 \ L_{\odot}$ and
the mass of the hot molecular gas $\rm \sim 9 \times 10^{-4} \
M_{\odot}$ (without any correction for extinction).

\end{abstract}



\bigskip

%
%
\section{Introduction}
\label{sect:intro} 
The Lagoon Nebula, Messier 8, is an H{\small II} region centred on the
stellar cluster NGC 6523. It is embedded within a molecular cloud
which extends to the young ($\sim 2 \times 10^6$ years old) star
cluster NGC\, 6530, $10'$ to its east (Lada et al.\ 1976). Star
formation is presumed to have proceeded from NGC\,6530 and is now
active in M8. Within M8's core lies the O7{\small V} star Herschel 36
(Woolf 1961), which has created a blister-type H{\small II} region,
the Hourglass.  This visually distinctive nebula is extended $15''$ EW
and $30''$ NS, and lies $15''$ E of Herschel 36.  The Hourglass is
embedded within an extended H{\small II} region, $\sim 3'$ in extent,
which is ionized by two O stars, HD\,165052 and 9\,Sgr (Woolf 1961,
Lada et al.\ 1976).  Nearby to Herschel 36 are a number of obscured
sources, first observed in the near--IR by Allen (1986). Woodward et
al.\ (1990) have designated them KS1 to KS5\@. Together, they may form
a cluster of hot stars, analogous to the Trapezium in the Orion Nebula
(M42), where $\rm \theta^1_c \ Ori$ is the dominant member. In the
mid--IR MSX\footnote{see http://irsa.ipac.caltech.edu/ for details.}
sky survey, this region appears as bright, extended source, with a
21\micron\ continuum flux of 960\,Jy. This source is also prominent in
sub--mm (Tothill 1999) and mm (Richter, Stecklum \& Launhardt 1998)
wavelength continuum. Nearby to Herschel 36, narrow band optical
imaging with the HST reveals an object reminiscent of the proplyds in
the Orion Nebula---a star within a bow-shock arc, whose apex is
pointed towards Herschel 36 (Stecklum et al.\ 1998). It is presumed to
be an externally ionized circumstellar disk.  These authors also
estimate the distance to M8 to be 1.8\,kpc, based on the association
with NGC\,6530.

Of particular interest to this paper are the observations by White et
al.\ (1997) of intense CO line emission from M8.  They found the peak
CO J=3--2 intensity to be over 100\,K, making it the second brightest
CO line source known. The CO J=3--2 and 4--3 lines were mapped and
found to have a loose, bipolar structure, extending NW--SE from
Herschel 36.  Taking the broad CO line profiles (extending over 20
\kms), and the presence of a jet-like object extending $0.5''$ SE of
Herschel 36 in HST images (Stecklum et al.\ 1995), it seems likely
that there is a molecular outflow in the core of M8\@.  If so, it
would be expected for there to be molecular hydrogen line emission as
well, both shocked (from the deceleration of the outflow by the
ambient cloud) and fluorescent (excited by far--UV photons from
Herschel 36).  However, White et al.\ (1997) also searched for the
near--IR H$_2$ v=1--0 S(1) line in M8, but did not detect it.  In this
paper we report more sensitive observations for H$_2$ emission from
M8, and find that there is indeed excited H$_2$ line emission from
this source.

\section{Observations and Data Reduction}
\label{sect:obs}
The Hourglass region of the Lagoon Nebula, Messier 8, was imaged on
1997, July 22, using the IRIS 1--2.5\micron\ camera, in conjunction
with the University of New South Wales Infrared Fabry-P\'{e}rot etalon
(UNSWIRF, Ryder et al.\ 1998), on the 3.9-m Anglo Australian Telescope
(AAT). The UNSWIRF etalon has a FWHM spectral resolution of $\rm \sim
75 \ km \ s^{-1}$, a pixel size of $0.77''$ and a $100''$ circular
field of view. It is scanned through a spectral line of interest in
order to obtain an emission line image of a source, with minimal
contamination from any continuum radiation present.  It also permits
limited kinematic information to be obtained, through the emission
velocity of the line centre across the field of view.

The H$_2$ 1--0 S(1) (2.1218\micron) line was observed.  Following a
rapid scan to determine the plate spacing for the line centre, three
on-line plate spacings were used for imaging, spaced by $\rm 39 \ km \
s^{-1}$, together with an off-line setting, $\rm 840 \ km \ s^{-1}$ to
the blue.  Integration times were two minutes per plate spacing, and
in addition, a sky frame, $300''$ N, was obtained immediately after
each source frame.

The star BS\,6748 (K=4.57 mags.) served as the flux standard, and was
imaged at each etalon spacing.  The absolute accuracy in flux
calibration using a Fabry-P\'{e}rot etalon is around 30\%. A diffused
dome lamp provided a flat field for each etalon spacing. An arc lamp
was scanned through a free spectral range in order to wavelength
calibrate the etalon response for each pixel of the array.

Data reduction was undertaken using a custom software package using
{\small IRAF}\footnote{Image Reduction and Analysis Facility (see
http://www.iraf.noao.edu).}.  Frames are linearised, flat-fielded using
a dome flat at the appropriate plate spacing, cleaned of bad pixels,
sky-subtracted, shifted to align stars in each frame, smoothed and the
off-line frame subtracted from each on-line frame (having been
appropriately scaled to minimise residuals from the subtraction
process). Stacking of the three on-line frames produces a data cube,
which is fitted, pixel-by-pixel, with the instrumental profile (a
Lorentzian), to yield a line image.  A coordinate frame was added
using the digitised sky survey\footnote{see
http://stdatu.stsci.edu/dss/.} to relate common stars between the
visible and 2.1\micron.

The line centre at each pixel position is obtained from the plate
spacing determined for the peak of the fitted profile.  The accuracy
of this procedure varies with position on the array, as well as with
the S/N the line is measured with. Where this is adequate, the
accuracy typically is $\rm \sim 10 \ km \ s^{-1}$.  Absolute line
emission velocities were determined with reference to the plate
spacing found for the peak of the H$_2$ emission in M17, which is
assumed to be at $\rm +21 \ km \ s^{-1}$ ($\rm V_{LSR}$). Note that,
as a consequence of the broad wings to the instrumental profile, the
line width cannot determined unless it is very much wider than the
spectral resolution (which it is not in M8).

\section{Results}
\label{sect:results}
\begin{figure}
\begin{center}
\begin{tabular}{c}
\psfig{figure=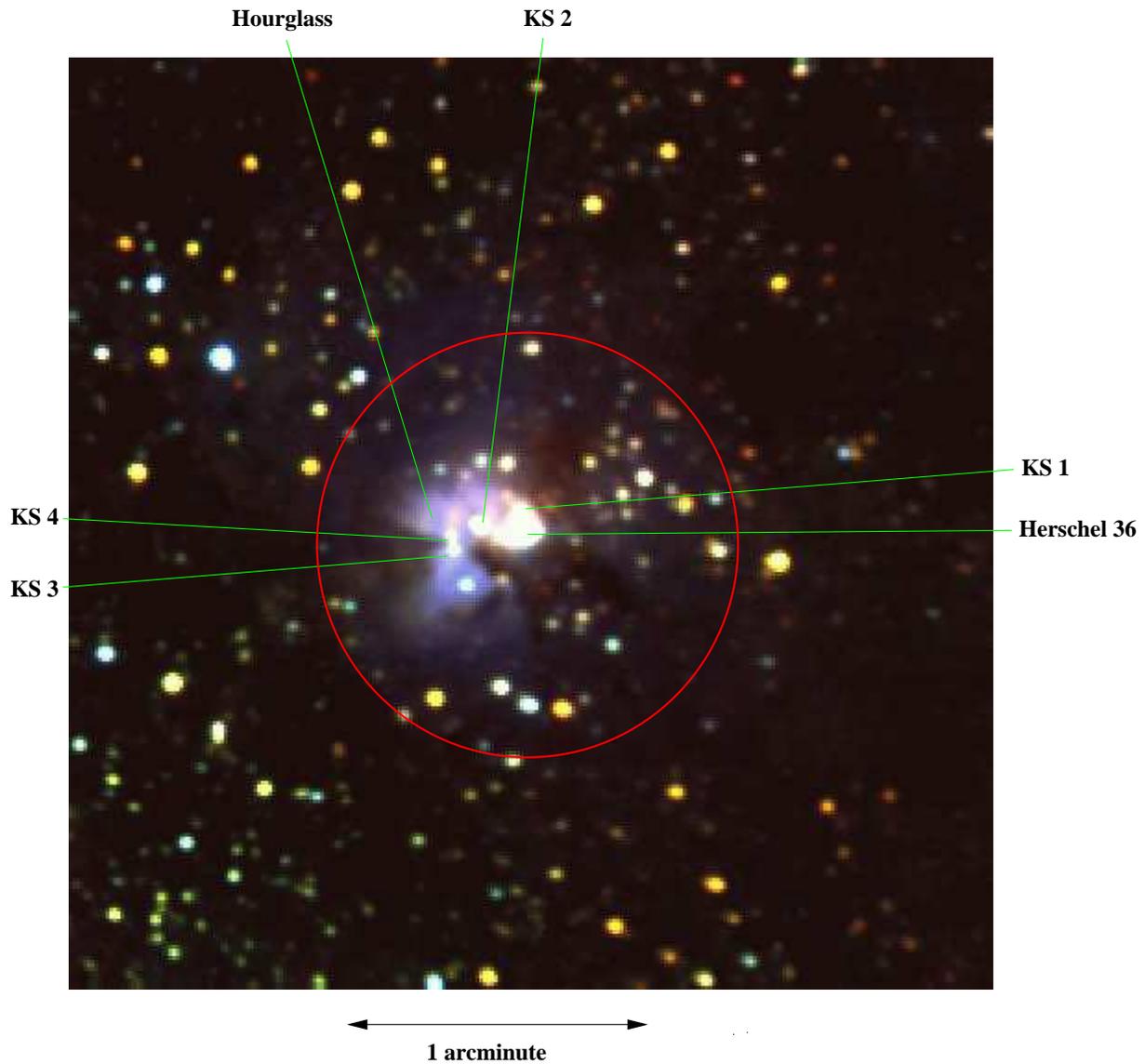,height=25cm,angle=0} 
\end{tabular}
\vspace*{-5cm}
\caption[] 
{Three-colour near-infrared image of Messier 8.  Blue denotes the J
(1.25\micron) band, green the H (1.65\micron) band and red the K
(2.2\micron) band.  The field imaged with UNSWIRF is indicated by the
red circle. Several features of interest are labelled; Herschel 36 the
O7{\small V} star powering the H{\small II} region emission, the
near--IR sources KS1 to KS4, and the Hourglass. Heavily reddened stars
appear red or yellow and foreground stars blue. The Hourglass is blue
because the near--IR emission is dominated by Paschen recombination
lines of hydrogen in the J--band. The scale bar shows an angular
distance of 1 arcminute.}
\label{fig:jhk} 
\end{center}
\end{figure}

The field surrounding the Hourglass is shown in the 3-colour near--IR
image in Fig.~\ref{fig:jhk}.  The J (1.25\micron), H (1.65\micron) and
K (2.2\micron) images were previously obtained by the author and David
Allen during the commissioning of IRIS in 1991. They have been
combined to produce a 3-colour image which simulates the appearance of
the nebula if our eyes were sensitive to the 1--2.5\micron\ wavelength
range.  Embedded and obscured stars are red or yellow and foreground
stars are blue.  The H{\small II} region of the Hourglass appears blue
as its near--IR emission is dominated by Paschen lines of hydrogen in
the J--band. The morphology of the Hourglass is the same in the three
bands, suggesting that this is its intrinsic shape, rather than being
determined by variable extinction in the optical (as suggested by
Woodward et al.\ 1986). In the figure, Herschel 36 and the Hourglass
are marked, as well as the four 2\micron\ sources KS1--KS4 (Woodward
et al.\ 1990).

\begin{figure}
\begin{center}
\begin{tabular}{c}
\psfig{figure=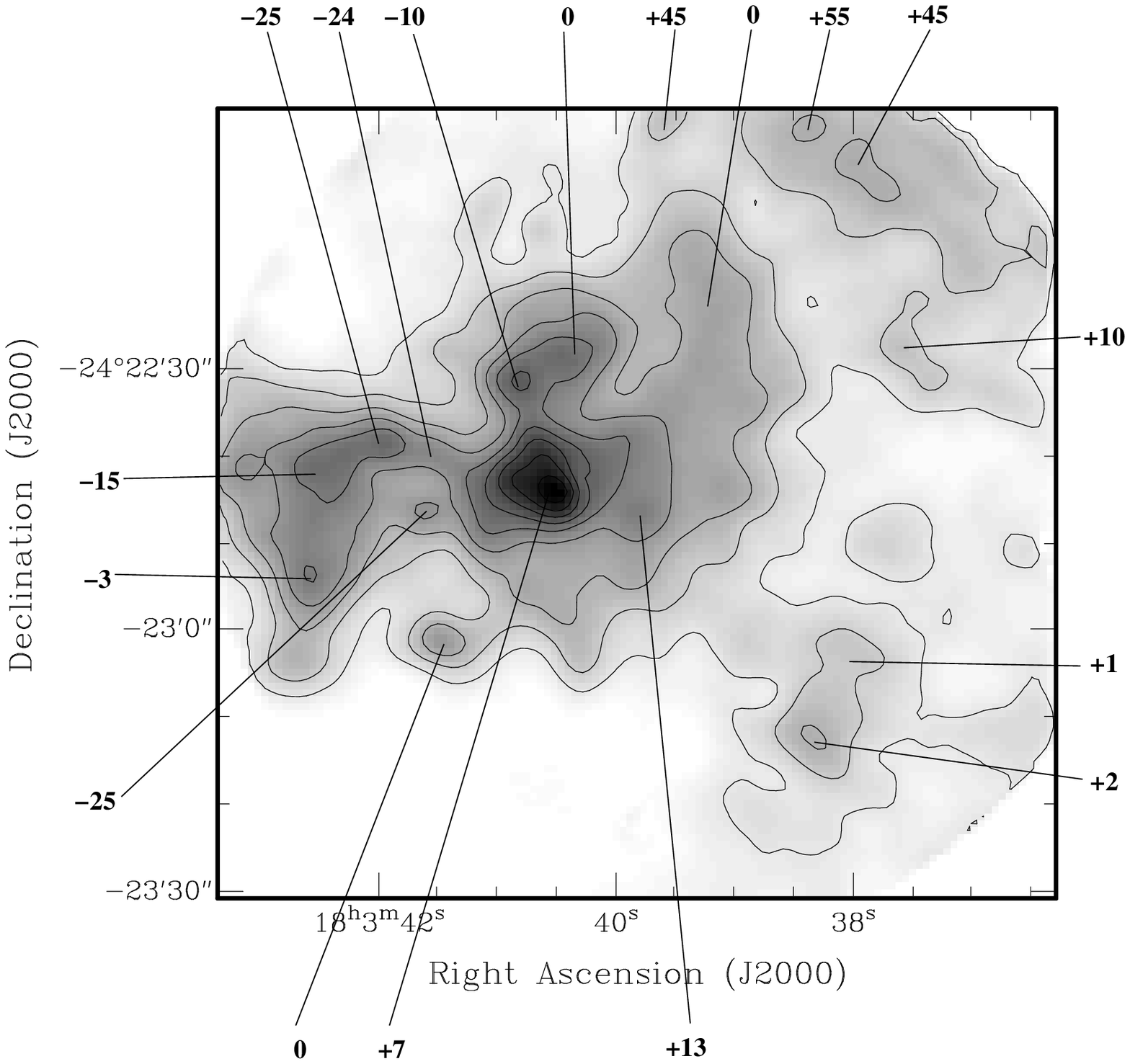,height=25cm,angle=0} 
\end{tabular}
\vspace*{-5cm}
\caption[] 
{Image of the molecular hydrogen v=1--0 S(1) line emission at
2.12\micron\ in Messier 8, overlaid with contours of the line emission.
Contours start at, and are in steps of, $\rm 8.5 \times 10^{-16} \
erg \ s^{-1} \ cm^{-2} \ arcsec^{-2}$.  Also indicated are the H$_2$
line centre velocities, in $\rm  \ km \ s^{-1}$ ($\rm V_{LSR}$), at
selected locations.}
\label{fig:h2} 
\end{center}
\end{figure}

The molecular hydrogen v=1--0 (1) line emission image of M8 is shown
in Fig.~\ref{fig:h2}, overlaid with contours of the line flux.  The
emission is clumped and the morphology broadly bipolar, centred about
Hershel 36. Also shown are the line centre velocities for selected
locations, obtained from fitting the instrumental profile to the three
on-line images.  While the accuracy of individual velocities is no
better than $\rm \sim 10 \ km \ s^{-1}$, they do indicate that there
is a velocity gradient of $\rm \sim 70 \ km \ s^{-1}$ across the
field, extending from $\rm \sim -25 \ km \ s^{-1}$ to the east of
Herschel 36, to $\rm \sim +45 \ km \ s^{-1}$ to its north-west.

\begin{figure}
\begin{center}
\begin{tabular}{c}
\psfig{figure=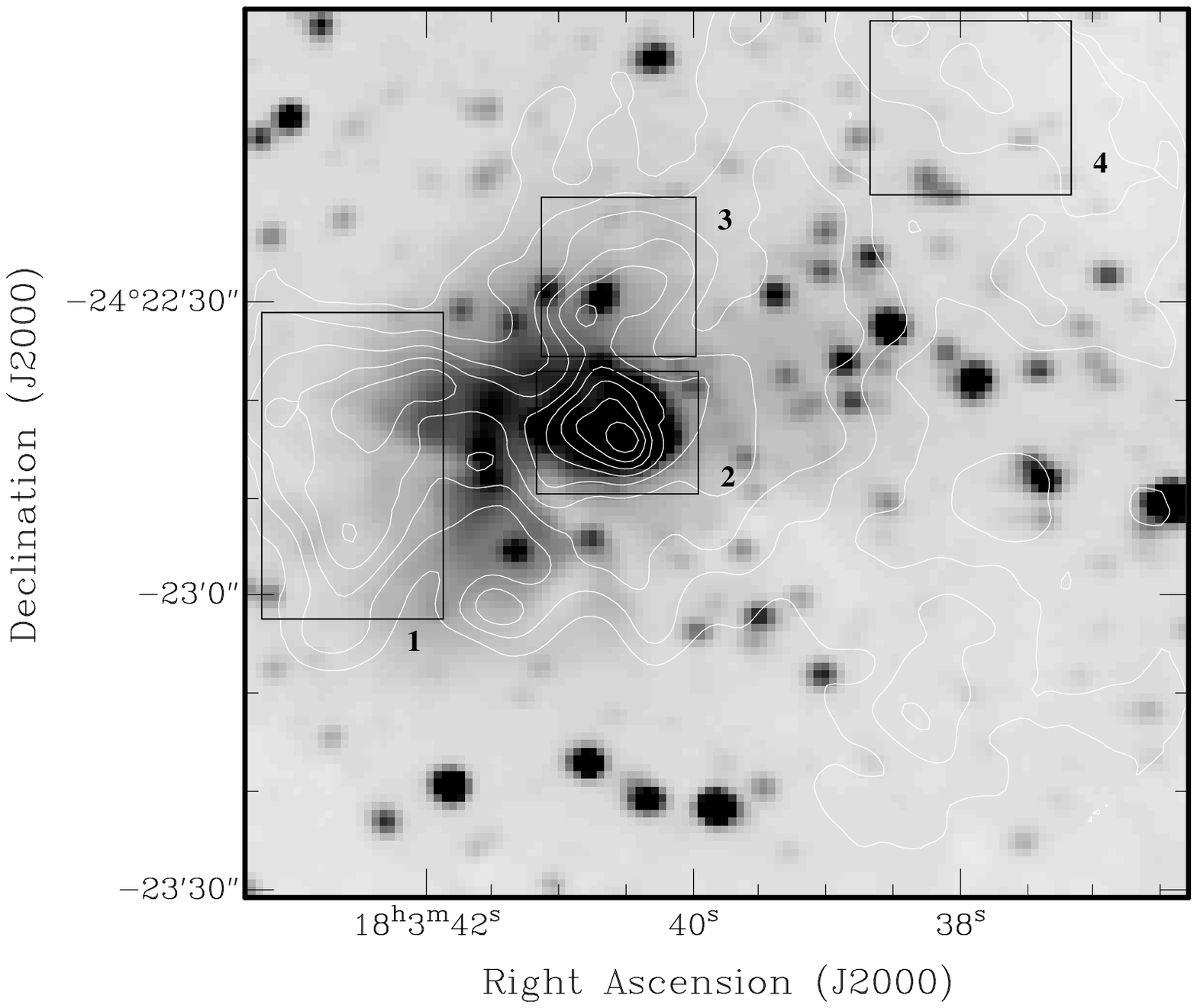,height=25cm,angle=0} 
\end{tabular}
\vspace*{-5cm}
\caption[] 
{Image of the K--band (2.2\micron) continuum emission from M8, overlaid
with contours of the H$_2$ 1--0 S(1) 2.12\micron\ line emission. Contours
are as in Fig.~\ref{fig:h2}.  The numbered boxes refer to the
apertures in Table~\ref{tab:fluxes}.}
\label{fig:kandh2} 
\end{center}
\end{figure}

The 2.12\micron\ H$_2$ line emission is compared to the 2.2\micron\
K--band continuum emission in Fig.~\ref{fig:kandh2}.  It can be seen
that the H$_2$ emission peaks close to the continuum peak, Herschel
36.  The Hourglass, however, appears as a cavity in the H$_2$
distribution.  H$_2$ line fluxes have been determined for the
apertures indicated in this figure, and are listed in
Table~\ref{tab:fluxes}, together with the positions of the emission
peaks.  The total 1--0 S(1) line emission from the region is $\rm 1.0
\times 10^{-11} \ erg \ s^{-1} \ cm^{-2}$ and the peak intensity is $\rm
8.2 \times 10^{-15} \ erg \ s^{-1} \ cm^{-2} \ arcsec^{-2}$. At an
assumed distance of 1.8\,kpc for M8, these figures are equivalent to a
luminosity in the 1--0 S(1) line of $\rm 1.1 \ L_{\odot}$, and to a
peak column density in the (v,J) = (1,3) level of the H$_2$ molecule
of $\rm 1.4 \times 10^{16} \ cm^{-2}$ (without any correction applied
for extinction to the emitting region).

\begin{table}
\caption[]{Molecular Hydrogen Line Fluxes in M8}
\label{tab:fluxes}
\begin{center}
\begin{tabular}{cccccc}
\hline
Region$^1$ & RA$^2$ & Dec$^2$ & Aperture$^3$ & 1--0 S(1) Flux$^4$ 
& 1--0 S(1) Peak Intensity$^5$  \\ 
& $18^h 03^m$ & $-24^{\circ} 22'$ & arcsec & 
{$\rm \times 10^{-13} \ erg \ s^{-1} \ cm^{-2}$} & 
{$\rm \times 10^{-15} \ erg \ s^{-1} \ cm^{-2} \ arcsec^{-2}$} \\
\hline
1 & $42.5^s$ & $22''$ & $22 \times 32$ & $17.6 \pm 0.1$ & $4.8 \pm 0.5$ \\
2 & $40.5^s$ & $44''$ & $17 \times 13$ & $11.1 \pm 0.1$ & $8.2 \pm 0.5$ \\
3 & $40.8^s$ & $31''$ & $17 \times 17$ & $9.2  \pm 0.1$ & $5.3 \pm 0.5$ \\
4 & $38.4^s$ & $02''$ & $20 \times 22$ & $7.8  \pm 0.1$ & $2.7 \pm 0.5$ \\
Total$^6$ & $40.5^s$ & $44''$ & $89 \times 79$ & $102 \pm 3$ & $8.2 \pm 0.5$ \\  
\hline
\end{tabular}
\end{center}
$^1$ Aperture, as shown in Fig.~\ref{fig:kandh2}. \\
$^2$ Position of peak emission within the aperture, in J2000. \\
$^3$ Size of aperture. \\
$^4$ Total H$_2$ 1--0 S(1) line emission from aperture. \\
$^5$ Peak emission intensity within aperture. \\
$^6$ Integrated fluxes for the entire field. \\
\end{table}

\begin{figure}
\begin{center}
\begin{tabular}{c}
\psfig{figure=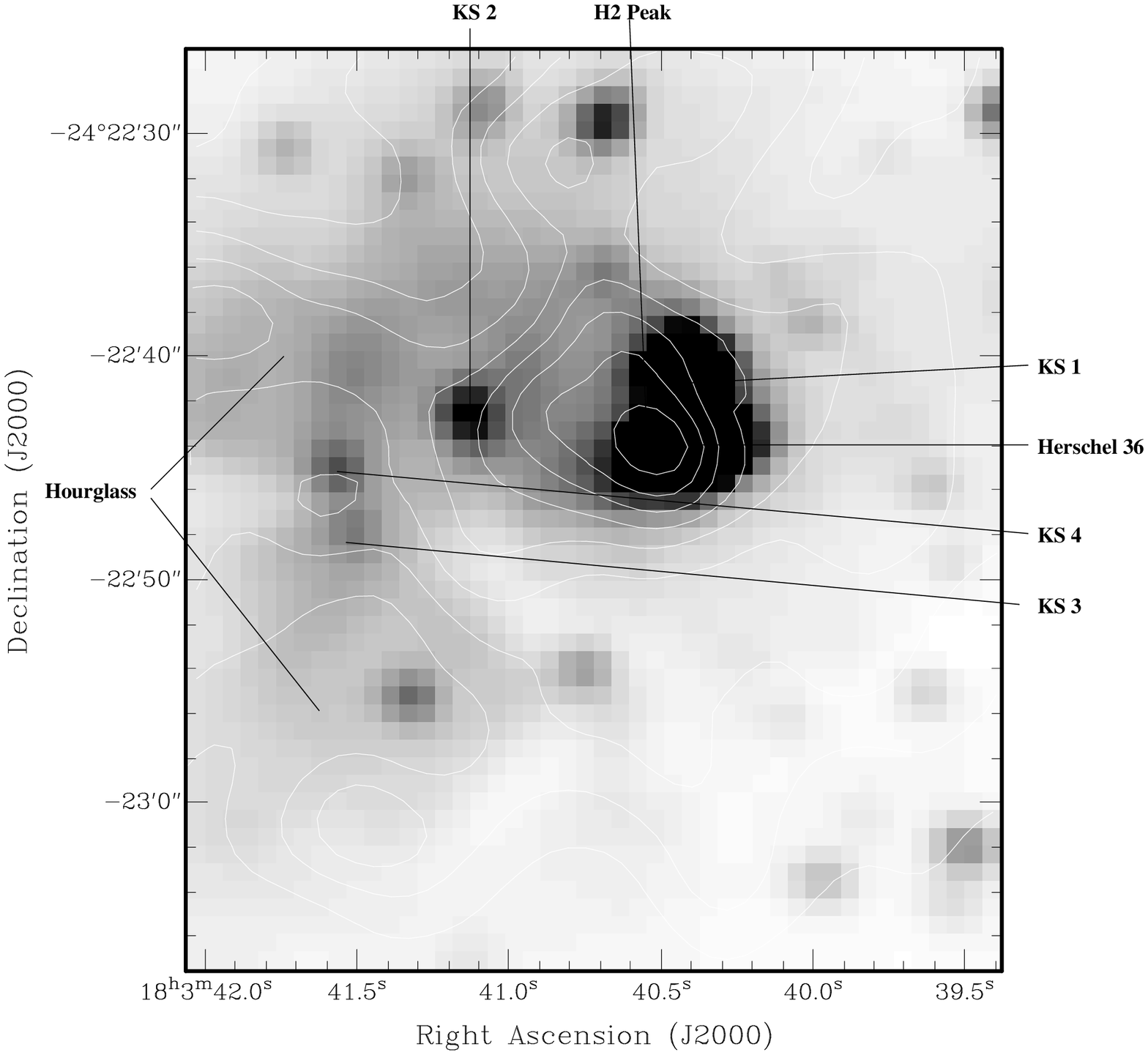,height=15cm,angle=-90} 
\end{tabular}
\vspace*{-1cm}
\caption[] 
{Enlarged image of the central region of M8, showing the K--band
continuum, overlaid by contours of the H$_2$ 1--0 S(1) line emission.
Contour levels are as in Figs.~\ref{fig:h2} and \ref{fig:kandh2}. The
separation between the H$_2$ peak and Herschel 36, the exciting star
for the Hourglass, is apparent. These features are labelled, together
with the infrared sources KS1--KS4.}
\label{fig:kandh2centre} 
\end{center}
\end{figure}

An enlarged image of the H$_2$ line emission, overlaid on the K--band
continuum, is shown in Fig.~\ref{fig:kandh2centre}.  It is apparent
that the H$_2$ emission does not, in fact, peak at the location of
Herschel 36. The H$_2$ emission peaks approximately (1.5''E, 0.5''S)
from Herschel 36, roughly half way between it and the proplyd (2.5''E,
1.5''S) of Herschel 36.

\section{Discussion}
\subsection{H$_2$ Morphology and Kinematics}
Given the intensity of the CO line emission measured in M8 by White et
al.\ (1997), it was a considerable surprise when the same authors
failed to detect any H$_2$ line emission from the source.  M8 is the
second most intense CO source detected, and clearly a site of active
massive star formation.  White et al.\ imaged M8 through a narrow band
(1\% width) filter centred on the H$_2$ 1--0 S(1) line, and used a
broad band 2.2\micron\ image to subtract the continuum from it.  They
failed to detect any H$_2$ line emission, and placed an upper limit of
$\rm \leq 3.4 \times 10^{-4} \ erg \ s^{-1} \ cm^{-2} \ sr^{-1}$ on
the line flux.  This is equivalent to $\rm \leq 8 \times 10^{-15} \
erg \ s^{-1} \ cm^{-2} \ arcsec^{-2}$, equal to the peak intensity we
actually measured in M8\@.  Thus, the upper limit of White et al.\
(1997) is consistent with our detection of H$_2$ line emission.  It
does, however, highlight the difficulty of measuring line emission
through narrow band filters in the presence of a strong continuum, and
the advantages of using a Fabry-P\'{e}rot etalon for such
measurements.  The $1 \sigma$ detection limit we achieved, with just 6
minutes on-line integration, is $\rm 3 \times 10^{-16} \ erg \ s^{-1}
\ cm^{-2} \ arcsec^{-2}$.

Comparison with the $^{12}$CO J=3--2 image obtained by White et al.\
(1997) (see their Fig.~2a, within the outermost white contour, at a
level of $\rm 230 \ K \ km \ s^{-1}$) shows that the morphology is
very similar to the H$_2$ 1--0 S(1) line, given the difference in
resolution of the two images (the CO 3--2 image was obtained through a
$14''$ beam).  The line emission is peaked around Herschel 36, and
extends in two lobes in a NW--SE, $\sim 45''$ from the centre.  The
extension to the SW is also seen in both line images. This corresponds
to a region of increased extinction seen in optical images.

The CO 3--2 line also displays broad line profiles, extending from +2
to +22 \kms\ ($\rm V_{LSR}$), and peaking at +10 \kms\ near Herschel
36.  Profiles in the NW lobe tend to extend somewhat further to the
red than they do in the SE lobe.  Similar behaviour is displayed by
the H$_2$ line centre velocities, though the velocity extent is
considerably larger than for the CO line.  The H$_2$ velocity ranges
from $\sim -25$ \kms\ in the SE lobe, to +7
\kms\ at the emission peak, to $\sim +45$ \kms\ in the NW lobe.

\subsection{Shock Excitation}
High spatial resolution imaging in the optical lines of H$\alpha$ and
[SII] with the HST, and with adaptive optics in the K and L band
near--IR continuum (Stecklum et al.\ 1995) shows an ionized, jet-like
feature, extending $\sim 0.5''$ from Herschel 36 to the SE (the K and
L band emission is presumed to dominated by the 2.16\micron\
Br$\gamma$ and 4.05\micron\ Br$\alpha$ hydrogen recombination lines).
Taken with the morphology and velocity extent of both the CO and H$_2$
line emission, this suggests that the H$_2$ line is shock-excited.  It
would appear to be the result of an interaction of a wind or jet from
Herschel 36 with the ambient molecular cloud. The H$_2$ morphology
then represents the location where a loosely collimated bipolar flow
from Herschel 36, surrounding an ionized jet, runs into a molecular
cloud.  The larger velocities seen in H$_2$ than CO represent the
shock speeds where the outflow impacts the molecular cloud.  The flow
then cools and decelerates, and this compressed gas is seen in the
lower-velocity CO lines.

Presuming the H$_2$ line emission is dominated by shocks, it is
possible to estimate the total H$_2$ line luminosity and the mass of
hot, post-shock gas.  In molecular shocks, the total H$_2$ line
luminosity is typically $\sim 15$ times the 1--0 S(1) line flux, and
the fraction of hot H$_2$ that is in the (v,J)=(1,3) state of the
molecule is $\sim 2\%$ (e.g.\ see Burton 1992).  The total H$_2$ line
luminosity from the source is then $\rm \sim 16 \ L_{\odot}$, and the
peak column density of hot H$_2$ is $\rm 6.1 \times 10^{17} \ cm^{-2}$
(i.e.\ at the emission peak).  The total mass of hot H$_2$ in the
source is $\rm \sim 9
\times 10^{-4} \ M_{\odot}$.  These figures should be corrected for an 
unknown amount of extinction to the emitting region.  This cannot be
obtained from our data, so we will use the estimates of $\rm A_v \sim
5$\,mags.\ towards Herschel 36 and the proplyd, made by Stecklum et
al.\ (1998). This is equivalent to $\sim 0.5$\,mags.\ at 2.1\micron.
If this is representative of the extinction to the H$_2$ emitting
region, then the extinction-corrected luminosity, peak column density
and mass of hot H$_2$ are $\rm \sim 25 \ L_{\odot}$, $\rm \sim 10^{18}
\ cm^{-2}$ and $\rm \sim 1.5 \times 10^{-3} \ M_{\odot}$,
respectively.

These figures can be compared to the total amount of molecular gas and
its peak column density, estimated by White et al. (1997) from the CO
lines.  They determined these to be $\rm \sim 31 \ M_{\odot}$ and $\rm
1.3 \times 10^{23} \ cm^{-2}$, respectively.  These are very much
greater than the quantities determined for the hot molecular gas, as
expected if the hot material comprises just a thin skin of material in
front of cooled, compressed gas.  The cooling time for hot H$_2$, $\rm
\tau_{cool}$, is $\sim 1$\,year, which therefore suggests that the outflow must
have existed for $\sim 10^{4-5}$\,years.

The mechanical luminosity of the outflow can be estimated as
\begin{equation}
{\rm L_{mech} \sim 0.5 \ M_{hot} V^2 / \tau_{cool}},
\end{equation}
where $\rm M_{hot}$ is the mass of hot H$_2$ and V is the average
speed of the shock wave.  Taking this to be 35\,\kms\ (half the range
in velocity of the H$_2$ line emission) yields $\rm L_{mech} \sim 300
\ L_{\odot}$ (extinction corrected), an order of magnitude 
greater than the H$_2$ line luminosity.  This is obviously a crude
estimate, but it does indicate that the H$_2$ lines are a significant
contributor, and possibly the dominant one (given the assumptions
made), to the cooling of the gas behind the shock front.

\subsection{Fluorescence}
Despite our conclusion that the H$_2$ line emission is dominated by
shocks, there may still be a significant contribution by far--UV
fluorescence to parts of it.  The Hourglass is an H{\small II} region,
excited by the UV radiation from Herschel 36.  The gap in the H$_2$
emission around the Hourglass suggests a physical connection between
the two, for instance the Hourglass may be a cavity embedded within
the outflow.  The H$_2$ line velocity in the cavity covering the
Hourglass is approximately constant, taking the value $\sim
-25$\,\kms\ ($\rm V_{LSR}$).  A constant velocity would be expected
for fluorescent emission from a molecular cloud with small internal
motions (e.g.\ see Burton et al.\ 1990).  The velocity measured then
indicates the $\rm V_{LSR}$ velocity of the surface of the
photodissociation region.  This velocity, $-25$\,\kms, is at the
negative limit of the 50 \kms\ range for the ionized gas in the
Hourglass, as measured through [OIII] 5007\AA\ line emission
(Chakraborty \& Anandarao 1999).  Furthermore, Woodward et al.\ (1986)
have detected 3.28\micron\ PAH emission from the region around the
Hourglass.  This is excited by the same far--UV photons that can
fluorescently excite H$_2$ (e.g.\ see the PAH emission in NGC\,6334;
Burton et al.\ 2000).  Therefore, it seems likely that some of the
H$_2$ emission, especially in the vicinity of the Hourglass, is
fluorescently excited.  However, with our data, it is not possible to
estimate what fraction this might be.

\section{Conclusions}
Strong near--IR molecular hydrogen line emission is produced from
around the Hourglass in the M8 star forming region.  The peak H$_2$
brightness is at the detection limit of a previous attempt to measure
H$_2$ line emission from the source, thus explaining the reported
non-detection.  The H$_2$ is emitted from an extended, roughly bipolar
region, centred on the powering source for the Hourglass, Herschel
36. It is orientated along a NW--SE direction.  The morphology is
similar to that of the CO J=3--2 distribution in M8.  Taken with the
$\sim 70$\,\kms\ variation in the H$_2$ line centre across the region,
and the 20\,\kms\ width of the CO profiles, this suggests that the
H$_2$ emission is shock-excited, when a bipolar outflow, originating
from Herschel 36, impacts the surrounding molecular cloud.  The total
H$_2$ line luminosity is $\sim
\rm 16 \ L_{\odot}$ (not corrected for any extinction).  The H$_2$ lines
provide a significant means of ridding the source of the mechanical
energy in the outflow, though we cannot yet determine whether H$_2$
line emission is the dominant coolant in the shocked gas.

To examine the excitation mechanism more closely, in particular to
determine whether there may also be fluorescent H$_2$ emission, it
would be necessary to measure higher excitation lines of the molecule,
from $\rm v \ge 2$.  This could be done either by long-slit
spectroscopy, or through further Fabry-P\'{e}rot imaging, with the
waveplates tuned to appropriate lines.  To improve the determination
of the flow energetics it will be necessary to use greater spectral
resolution and resolve the line profiles, in order to measure the
momentum and kinetic energy distribution of the hot molecular gas
across the source.

\section*{Acknowledgements}


Several people have provided considerable assistance with the work
presented here.  I particularly wish to thank Michael Ashley, Lori
Allen, Jung-Kyu Lee, Stuart Ryder and John Storey for their help with
the observations.  Christine Carmody also helped with the reduction of
the data.  The staff of the Anglo Australian Telescope provided great
assistance with the installation and operation of UNSWIRF.

\section*{References}






\reference Allen, D.A., 1986, MNRAS, 319, 35P.

\reference Burton, M.G., 1992, Aust. J. Phys., 45, 463.

\reference Burton, M.G., Ashley, M.C.B., Marks, R.D., Schinckel,  A.E., 
Storey, J.W.V., Fowler, A., Merrill, M., Sharp, N., Gatley, I.,
Harper, D.A., Loewenstein, R.F., Mrozek, F., Jackson, J. \& Kraemer,
K., 2000, ApJ, 542, 359.

\reference Burton, M.G., Geballe, T.R., Brand, P.W.J.L. \& Moorhouse, A.,
1990, ApJ, 352, 625.


\reference Chakraborty, A. \& Anandarao, B.G., 1999, A\&A, 346, 947.

\reference Lada, C.J., Gull, T.R., Gottlieb, C.A. \& Gottlieb, E.W., 1976, 
ApJ, 203, 159.

\reference Richter, S., Stecklum, B. \& Launhardt, R., 1998, Abstracts of 
contributed talks and posters presented at the annual scientific
meeting of the Astronomische Gesellschaft, Heidelberg, September
14-19, 1998.

\reference Ryder, S.D., Allen, L.E., Burton, M.G., Ashley, M.C.B. \& 
Storey, J.W.V., 1998, PASA, 15, 228.

\reference Stecklum, B., Henning, T., Eckart, A., Howell, R.R. \& Hoare, M., 
1995, ApJ, L153.

\reference Stecklum, B., Henning, T., Feldt, M., Hayward, T.L., Hoare, M.G., 
Hofner, P. \& Richter, S., 1998, AJ, 115, 767.

\reference Tothill, N.F.H., 1999, PhD Thesis, Queen Mary and Westfield 
College, University of London.

\reference White, G.J., Tothill, N.F.H., Matthews, H.E., McCutcheon, W.H., 
Huldtgren, M. \& McCaughrean, M.J., 1997, A\&A, 323, 529.

\reference Woodward, C.E., Pipher, J.L., Helfer, H.L., Sharpless, S., 
Moneti, A., Kozikowski, A., Oliveri, M., Willner, S.P., Lacasse,
M.G. \& Herter, T., 1986, AJ, 91, 870.

\reference Woodward, C.E., Pipher, J.L., Helfer, H.L. \& Forrest, W.J., 
1990, ApJ, 365, 252.

\reference Woolf, N.J., 1961, PASP, 73, 206.

\vfill\eject 

\end{document}